\documentclass[aps,prb,groupedaddress,superscriptaddress,amsmath,amssymb,showpacs,preprint]{revtex4-1}

\usepackage{graphicx}
\usepackage{dcolumn} 
\usepackage{bm} 
\usepackage{epsfig}
\usepackage{subfigure}
\usepackage{color}

\begin{document}

\title{\color{black} Modelling the influence of photospheric turbulence on solar flare statistics}
\author{M. Mendoza} \email{Correspondence and requests for materials
  should be addressed to M. Mendoza (mmendoza@ethz.ch)} \affiliation{ETH
  Z\"urich, Computational Physics for Engineering Materials, Institute
  for Building Materials, Schafmattstrasse 6, HIF, CH-8093 Z\"urich,
  Switzerland}

\author{A. Kaydul} \affiliation{ETH Z\"urich, Computational Physics
  for Engineering Materials, Institute for Building Materials,
  Schafmattstrasse 6, HIF, CH-8093 Z\"urich, Switzerland}

\author{L. de Arcangelis} \affiliation{Department of Industrial and 
Information Engineering, Second University of Naples and CNISM, 81031 
Aversa (CE), Italy}

\author{J. S. Andrade Jr.} \affiliation{Departamento de F\'{\i}sica,
  Universidade Federal do Cear\'a, 60451-970 Fortaleza, Cear\'a,
  Brazil}

\author{H. J. Herrmann} \affiliation{ETH Z\"urich, Computational
  Physics for Engineering Materials, Institute for Building Materials,
  Schafmattstrasse 6, HIF, CH-8093 Z\"urich, Switzerland}
\affiliation{Departamento de F\'{\i}sica,
  Universidade Federal do Cear\'a, 60451-970 Fortaleza, Cear\'a,
  Brazil}

\date{\today}


\maketitle
\pagebreak

\maketitle

{\bf Solar flares stem from the reconnection of twisted magnetic field
  lines in the solar photosphere.  The
  energy and waiting time distributions of these events follow complex patterns that have been carefully considered in the past and that bear some
  resemblance with earthquakes and stockmarkets. Here we explore in detail the tangling motion of
  interacting flux tubes anchored in the plasma and the energy
  ejections resulting when they recombine. The mechanism for energy accumulation and release in the flow is reminiscent of self-organized criticality.
  From this model we {\color{black} suggest the origin for} two important and widely studied properties of solar flare statistics,  including the time-energy correlations. We first  propose that the scale-free energy distribution of solar
  flares is {\color{black} largely} due to the twist exerted by the
  vorticity of the turbulent photosphere. Second, the long-range
  temporal and time-energy correlations {\color{black} appear to} arise from the tube-tube interactions. {\color{black} The agreement with satellite measurements is encouraging}.}

\section*{Introduction}

Parker conjectured that solar flares are driven by the random
continuous motion of the footprints of the magnetic field in the
photospheric convection \cite{parker88, parker89}. This conjecture and
the experimental observations of power laws in the energy \cite{exp1,
  exp2, exp3, hudson91} and waiting time \cite{norandom1, intertime1,
  intertime2} distributions stimulated a new way of looking at
violent bursts. In particular, the fact that the energy distribution
is a power law, a property that flares share with diverse physical
phenomena\cite{stocks}, such as avalanches and earthquakes \cite{arcangelis1}, led to the formulation of flare occurrence models inspired
in self-organised criticality (SOC)\cite{avalanche, avalanche2, nicodemi,
  model3, georgoulis1998, vlahos2004, dimi2011, dimi2013, morales2009, morales2008self, kras2002}. Although these models reproduce the power-law behaviour in the distribution of flare energies, they predict a Poissonian
distribution waiting times, which implies that
flares result from an uncorrelated process, contrary to experimental
observations \cite{norandom1, intertime1, intertime2}. This point was
stressed by Boffetta {\it et al.}\cite{norandom1} who, by implementing
shell models for turbulence, reproduced the observational power-law
decay of the waiting time distribution. However, their exponents are
not universal, {\color{black} depending instead on the model parameters}.

Based on a different approach, the waiting time distribution can also be
reproduced in terms of a piecewise Poissonian process
\cite{wheatland1}. More recently, the existence of correlations
between flare energies and waiting times has also been investigated \cite{energyintertime1, intertime2, energyintertime2,
  lucilla2}. In particular, Lippiello {\it et al.}  \cite{lucilla2}
found that the observed time-energy correlations are not simply attributable to obscuration effects. Here the term obscuration indicates an observational limitation that can be at the origin of the incompleteness of the
catalogue, for example, the occurrence of a large flare can hide the
detection of smaller flares occurring nearby.

An approach more closely inspired in magnetic reconnection was adopted by
Hughes {\it et al.}\cite{model3}, who proposed a dynamical model of
solar flares as cascades of reconnecting magnetic loops, with multiple
loops that are randomly driven at their footprints and interact with
each other. Despite some discrepancy with experimental observations,
they showed a relation of the distribution of magnetic loops with a
scale-free network, which conceptually supports self-organised
criticality.

{\color{black} The formulation of a theoretical model able to reproduce both the flare statistics and the behaviour of time-energy correlations remains unsolved.} A complete model would require a fully developed realistic convection zone, a stratified atmosphere above it and the study of the interplay between magnetic fields and flows. Additionally, the model must be three-dimensional, including explicit resistivity in order to control the reconnection between the colliding magnetic
fields. At present, a numerical study of such a model is far out of reach due to the excessive requirement in computing time. On the other end, simplified
models considering magnetic reconnections based on a purely
statistical approach and neglecting completely the photospheric flow,
fail in reproducing a variety of observational data \cite{avalanche, avalanche2, nicodemi,model3, georgoulis1998, vlahos2004, dimi2011, dimi2013, morales2009, morales2008self, kras2002}. 

In our study, we present {\color{black} an approach that} represents a compromise between these two different scenarios. Here we introduce and study numerically a theoretical model of solar flare occurrence in terms of reconnection of magnetic flux tubes twisted by the photospheric turbulent flow, which reproduces satisfactorily the flare statistics and the behaviour of time-energy correlations. The motion of the tubes in the solar corona is mainly rooted in the photosphere, which is about $\sim 500$ km thick, corresponding to an extremely thin layer as compared to the solar radius.  Therefore, in the model proposed here we consider the photospheric flow as a two-dimensional turbulent system following Kolmogorov scaling.

\section*{Results}

\subsection*{Model description}

The turbulent fluid dynamics of the photosphere is simulated through a lattice Boltzmann model\cite{LBE1} on a square lattice of size $L$, with a forcing term that specifically reproduces the Kolmogorov energy 
spectrum regime (see Methods). 

\begin{figure}
  \centering
  \includegraphics[width=0.5\columnwidth]{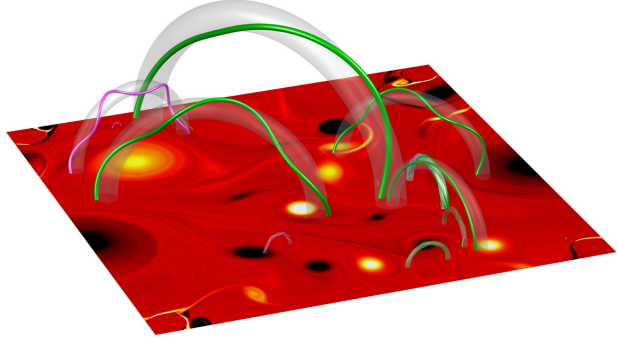}
  \caption{{\bf Configuration of magnetic tubes in the solar corona}.
    The photospheric plasma is in
    a turbulent state, where yellow (black) areas denote high (low)
    vorticity regions.  The green and pink lines represent the
    magnetic lines enclosed in the magnetic flux tubes
    (semi-transparent grey tori). The pink magnetic lines indicate a
    magnetic flux tube that has reached the critical twist and,
    therefore, is at the onset of reconnection, about to release its energy as a flare.}\label{fig1}
\end{figure}

Anchored in the photospheric flow, the footprints of the magnetic flux
tubes follow the local velocity field, and are twisted by the
vorticity. The magnetic lines are modelled as lines (see green and pink lines in Figure~\ref{fig1}) wrapped around the semi-circular flux tubes (see semi-transparent grey tori in Figure~\ref{fig1}), forming, when twisted, a spring-shaped bundle. This representation, which is conceptually consistent with previous realistic models for the kink instability
\cite{hood2}, leads to an explicit relation between the length of the
spring and the magnetic energy stored in the corresponding flux tube (see Eq. \eqref{energy_tube} in Methods).

Observational evidence supports the kink instability as the triggering
mechanism for magnetic reconnection, and consequently, for solar flare occurrence \cite{priest2007magnetic, expkink1, expkink2, kliem2004}. The kink instability is an ideal magnetohydrodynamical (MHD) instability, where magnetic reconnection is not a necessary ingredient, as a flux tube can destabilise by converting twist into writhe. However, in an active region (which is our region of interest), the kink instability will trigger a flare (due to the accumulated free energy), and therefore, we assume that each time a kink instability occurs in a magnetic tube, a flare is released. Accordingly, the kink instability of a flux tube occurs as soon as the intensity of its cumulative twist reaches a given critical value. In our model, the tube releases its total energy (vanishing from the simulation), and a new magnetic flux tube is inserted with the initial condition and located at a random
position inside the simulated zone (see Methods). The critical twist $\Phi_{\rm c}$, at which the magnetic reconnection occurs, can be obtained from stability
analysis. Several values have been proposed from numerical
simulations, theoretical models \cite{hood2} or deduced from
experimental observations \cite{expkink1, expkink2}, ranging from
$2\pi$ to $12\pi$, depending on the particular plasma conditions in
the corona.

Every time a magnetic flux tube reconnects and releases its energy, we
implement two possible scenarios: The reconnection
heats up the surrounded plasma increasing the local coronal pressure,
and therefore, increasing the critical twist of the surrounding
magnetic flux tubes \cite{hood2}. This causes a delay in the reconnection process, which is taken into account here by multiplying the
cumulative twist of neighbouring tubes by a positive factor
$\lambda_{\rm R} < 1$, while keeping the critical twist constant. This procedure implements tube-tube interactions and in our simulations we consider either random values or a constant value for $\lambda_{\rm R}$, obtaining the same results. Conversely, we also consider the case $\lambda_{\rm R} > 1$ which induces an avalanching process in the occurrence of solar flares. In this case, every time a magnetic flux tube releases its energy it increases the twist of the surrounded tubes, triggering new flares and generating a cascade of events. In both cases, the parameter $\lambda_{\rm R}$ has the important role of tuning the level of tube interactions, whereas non-interacting tubes correspond to $\lambda_{\rm R}=1$. We are therefore able to detect how interactions affect the flare waiting time distribution. In what follows, we show that tube-tube interaction is the most relevant ingredient to reproduce the correct temporal organisation and the
experimentally observed time energy correlations. We will also show
that changing the value of the critical twist within the range of the
experimentally meaningful values, does not affect the statistics of
solar flares. Our model takes into account the well-known SOC mechanism of slow energy storage by changing the flaring threshold of tubes that are neighbours to the reconnected tube. This mechanism is equivalent to tuning the critical tube twist, a procedure usually implemented in SOC models \cite{avalanche, avalanche2, nicodemi,model3, georgoulis1998, vlahos2004, dimi2011, dimi2013, morales2009, morales2008self, kras2002}. More precisely, by increasing the critical twist ($\lambda_{\rm R} < 1$) , the photospheric flow has more time to add energy to the respective magnetic flux tubes before they release their energy as a flare. For the case $\lambda_{\rm R} > 1$, the similarity with SOC models is more evident, since our model can generate a cascade of events by increasing the twist of surrounded magnetic flux tubes, each time a flare is released. Finally, we should mention that our approach does not account for a number of additional features of flare occurrence and flaring active regions \cite{forbes2006cme,aulanier2013}, e.g. systematic peculiar flows on top of the heavily suppressed quiet-Sun Kolmogorov flow velocity field, the presence of intense magnetic polarity inversion lines (PILs) in the photosphere, and slip-running magnetic reconnection in flares. However, the excellent agreement between numerical and observational data suggests that these features are not relevant to produce the observed statistical properties, even if they may be necessary for the complete understanding of solar flares.

\subsection*{Peak and Total Energy Distributions}

Following the described dynamics, the occurrence time
and energy released by each flare quantify the statistical
properties of our model and are compared with observational data.  We perform
extensive numerical simulations for different system sizes $L=128$,
$256$, $512$, $1024$, and $2048$. For each value of $L$, we first let
the fluid evolve without the magnetic flux tubes until it reaches the
turbulent regime at Reynolds numbers between $10^4$ and $10^5$.  The Reynolds number is computed by the relation $Re = u_{\rm rms} L/\nu$, where $u_{\rm rms}$ is the root mean square velocity, and $\nu$ the kinematic viscosity (values of the Reynolds number and kinematic viscosity for each simulation can be found in the Supplementary Table 1). The fully developed turbulence condition is established when the energy spectrum of the flow follows the classical power-law behaviour with Kolmogorov exponent $-5/3$ (see Supplementary Figure 1).  At this point, we add $N$ magnetic flux tubes at random positions and with random initial energies, and record the occurrence time and energy of flares. 
From previous studies of satellite data by the Solar and Heliospheric Observatory (SOHO), deviations from the Kolmogorov exponent have been observed depending on the activity of a solar region\cite{kolmogorov_MHD}. However, we show that the choice of the exponent of the spectrum of the turbulent flow does not change appreciably our results (see {\color{black} Supplementary Figure 2}).

\begin{figure}
  \centering
  \includegraphics[width=0.5\columnwidth]{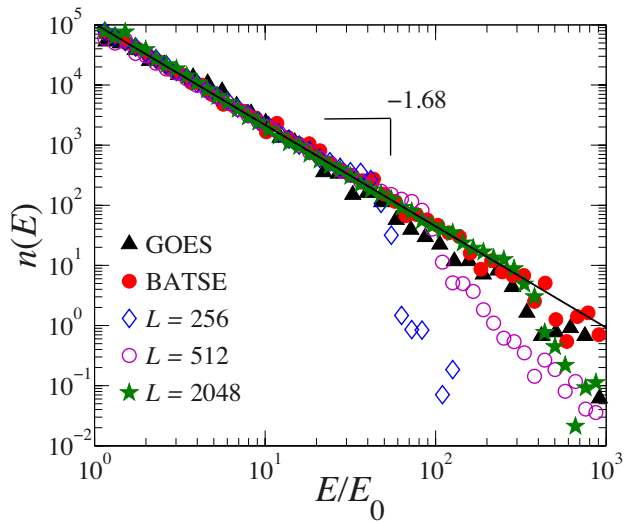}
  \caption{{\bf Solar flare energies exhibit scale-free
      behaviour}. The distribution of peak flare energies evaluated with our numerical model for different system sizes $L$ and $\lambda_{\rm R} < 1$ shows a power-law regime which increases with $L$. The solid line is a fit for the largest system size, $L = 2048$, providing an exponent of $\alpha = 1.68 \pm 0.02$.  Observational data from GOES and BATSE catalogues \cite{GOES, BATSE} exhibit the same scaling behaviour. Energies are expressed in units of a lower energy cutoff, which is $E_0 = 10^{-6}$ W m$^{-2}$ (GOES) and $E_0 = 0.5$ cmnts sec$^{-1}$ cm$^{-2}$ (BATSE) for observational data, and $E_0 = 10$ for the numerical catalogues. We have verified that different cutoff values do not affect the scaling behaviour.}\label{fig2}
\end{figure}

\begin{figure}
  \centering
  \includegraphics[width=\columnwidth]{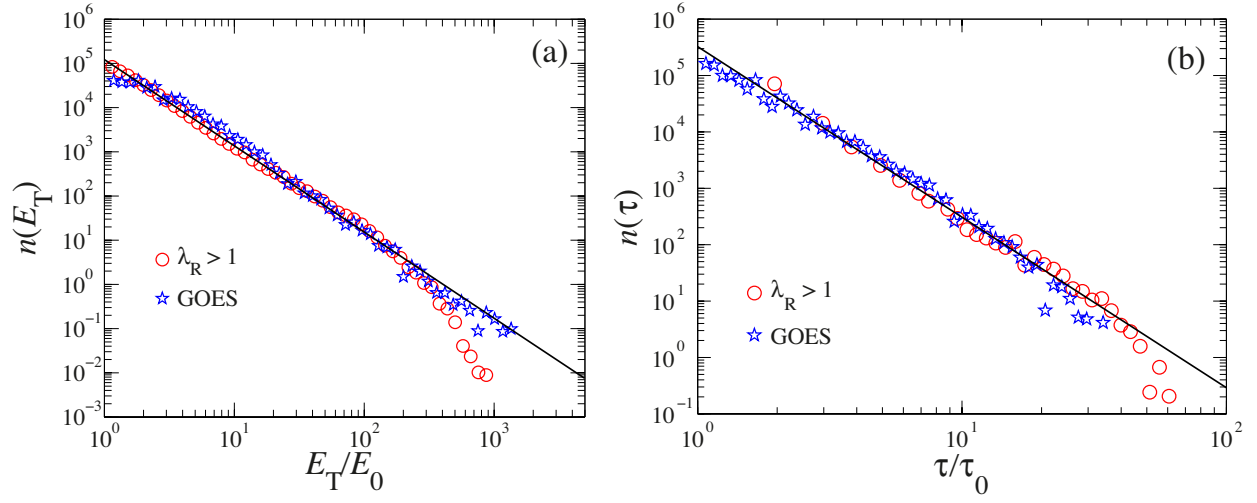}
  \caption{{\bf Duration of flares and total energy distributions}. The distributions of duration of flares and total flare energies are evaluated with our numerical model for $\lambda_{\rm R} > 1$, using a system size of $L=512$ and $N = 400$, and compared with observational data from the GOES catalogue \cite{GOES}. (a) Total energy distribution. The distribution exhibits a scale-free behaviour with an exponent of $-1.95 \pm 0.04$ (solid line). Energies are expressed in units of a lower energy cutoff, which is $E_0 = 6\times 10^{-7}$ J m$^{-2}$ and $E_0 = 11.4$ for the observational and numerical data, respectively. (b) Distribution of flare durations. Here $\tau_0 = 10^3$ s for GOES catalogue and $\tau_0 = 1$ for the numerical data, corresponding to the shortest time at which the power-law regime is observed. The solid line corresponds to an exponent of $-3.0 \pm 0.1$.  
  }\label{fig5tt}
\end{figure}

As shown in Figure~\ref{fig2}, the distribution of peak flare energies follows a typical power-law behaviour, $n(E) \propto E^{-\alpha}$, with
an exponent $\alpha=1.68 \pm 0.02$.  As expected, this scaling regime
extends up to a cutoff energy that gradually increases with system
size $L$. We compare our results with soft and hard X-ray data from
the GOES \cite{GOES} and the BATSE \cite{BATSE} catalogues,
respectively. From the GOES catalogue, only flare events of class C
and above (peak flux $E > E_{0}= 10^{-6}$ Watt m$^{-2}$) observed between $1992$
and $2013$ are considered, which leads to a total of $19703$ events,
covering nearly two solar cycles. In the case of the BATSE catalogue,
$7242$ events in the period between $1991$ and $2000$ were considered, with a peak flux larger than $E_{0}= 0.5$ counts sec$^{-1}$ cm$^{-2}$.
 The Kolmogorov-Smirnov test ensures that both samples, numerical and observational data, follow the same scaling behaviour with a $p$-value of $95 \%$ (confidence level of $99\%$). The excellent agreement between numerical and observational data shown
in Figure~\ref{fig2} confirms the validity of the theoretical
approach. The exponent of the power law for our numerical results is also in excellent agreement with previous experimental studies
\cite{energyintertime1, arcangelis1}. Notice that in our model for $\lambda_{\rm R} < 1$, flares are instantaneous events, therefore we cannot measure separately the peak energy and the total energy associated to each event, or else the energy of a flare is a peak flux energy. Different is the case $\lambda_{\rm R} > 1$, where avalanching is induced and events are therefore not instantaneous. In order to implement a unified procedure for models with different $\lambda_{\rm R}$, numerical data are compared to peak flux energies from experimental observations. For the case of $\lambda_{\rm R} > 1$, one can also study the total energy distribution and the duration of each event. In {\color{black} Figure~\ref{fig5tt}}, we can see that the numerical results are in very good agreement with observations, showing the correct duration and total energy distributions. The exponents for the total energy distribution and the duration of flares, $-1.95 \pm 0.04$ and $-3.0 \pm 0.1$, respectively, are also in good agreement with previous studies~\cite{georgoulis1998}.

{\color{black} Note that our model results in steeper distribution functions for the total energy than for the peak released energy of the modeled events. On the other hand, numerous observational works report the opposite, that is, clearly flatter distribution functions for the total energy. Theoretical works, at least those relying on SOC, seem also to predict analytically that total-energy distributions functions are flatter than those of the peak energy released.}

\subsection*{Waiting Time Distributions}

\begin{figure}
  \centering
  \includegraphics[width=0.5\columnwidth]{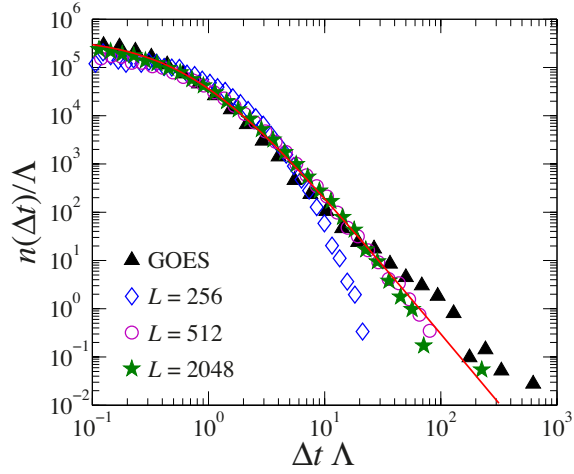}
  \caption{{\bf Solar flares exhibit a complex temporal organisation}.
    The waiting time distribution for solar flares is shown for
    different system sizes $L$ and compared to observational data for
    the GOES catalogue \cite{GOES}. The distributions indicate that the process is
    not a simple Poisson process. The solid line denotes the fit for
    the largest system size, $L = 2048$, using as a fitting function
    the analytical expression proposed in Ref.~\cite{fitting-time}. To
    better compare the different catalogues, waiting times are normalised
    by $\Lambda$, the average rate evaluated for each numerical and
    observational catalogue.}\label{fig4}
\end{figure}

We also investigate the statistical patterns of the waiting time, defined as the distribution of time delays between the end of an event and the beginning of the next one. As shown in Figure~\ref{fig4}, the
numerical results from our theoretical approach are also in very good
agreement with experiments.  The distribution is not a simple
exponential, suggesting that flare occurrence is not a purely
uncorrelated Poisson process. In order to closely compare the
different numerical and observational catalogues, we have rescaled the
waiting times, $\Delta t$, by the average event rate in each catalogue
\cite{arcangelis2}, i.e., by the inverse of the average waiting time,
$\Lambda = N_{\rm e}/(t_{\rm max}-t_{\rm min})$,  where $N_{\rm e}$ is the number of events in the respective catalogue.  Here $t_{\rm max}$ and $t_{\rm min}$ are
the times at which the last and first events in the catalogue occur,
respectively. As shown in Figure~\ref{fig4}, rescaled distributions for
numerical and observational data collapse onto a universal curve well
fitted by \cite{fitting-time}, $n(\Delta t \Lambda)/\Lambda = a/(1 + b
\Delta t \Lambda )^{\alpha_{\rm t}}$, where $a$ and $b$ are constants, and
$\alpha_{\rm t}=2.8\pm 0.2$ denotes the exponent of the power-law regime
of the distribution for large waiting times. This result is reasonable when compared with previously reported observational values, $\alpha_{\rm t} =
2.16 \pm 0.05$ \cite{fitting-time} and $\alpha_{\rm t} = 2.4 \pm 0.1$
\cite{norandom1}.  We have also performed the Kolmogorov-Smirnov test, finding that both samples, numerical and observational data, come from the same distribution with a $p$-value of $92 \%$ (confidence level of $99\%$).

\begin{figure}
  \centering
  \includegraphics[width=\columnwidth]{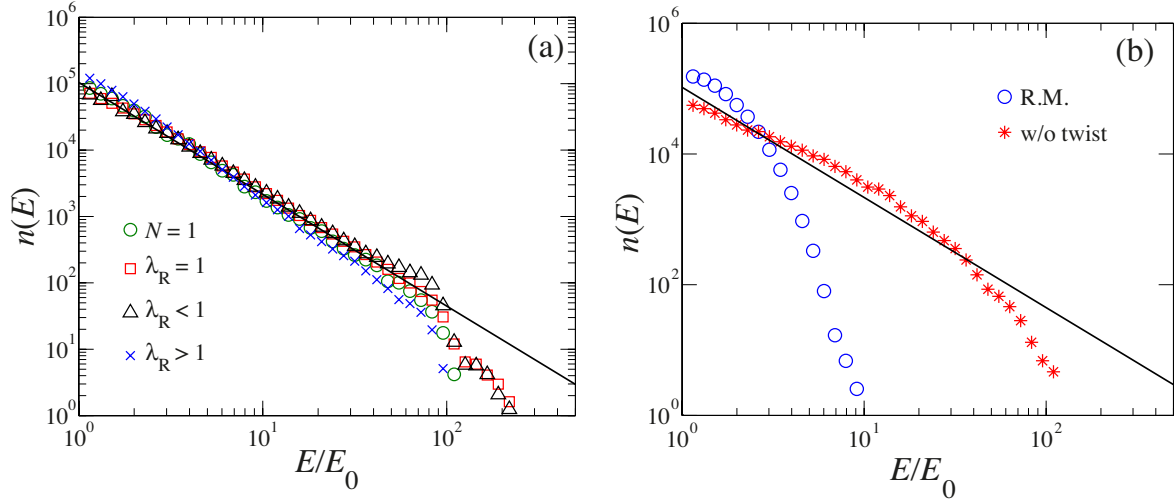}
  \caption{{\bf Physical ingredients leading to scale-free behaviour
      for flare energies}. Different models are compared to extract
    the crucial ingredients of the power law behaviour measured for
    the energy distribution. The system size is $L = 512$ for all
    curves. (a) Flare peak energy distribution for simulations
    where a single flux tube ($N = 1$) evolves in the turbulent flow
    and reconnects at twisting threshold. {\color{black} The power-law behaviour is
    recovered with exponent $\alpha = 1.68 \pm 0.02$ (solid line)}. The same
    behaviour is recovered when changing interactions: from
    interacting tubes ($\lambda_{\rm R} < 1$ and $\lambda_{\rm R} > 1$) to non-interacting ($\lambda_{\rm R} = 1$) tubes. (b) Solar flare peak energy distribution for the evolution
    of $N = 400$ interacting tubes ($\lambda_{\rm R} < 1$) in the case tube
    footprints diffuse by random motion (denoted by ``R.M.''). The
    distribution in this case does not show a power-law
    regime. For the same system size, we let the tubes evolve
    according to the turbulent flow but reconnection occurs only
    through the interaction of footprints or tube crossing (denoted by
    ``w/o twist''), i.e., twisting of magnetic tubes is not
    considered. Also in this case a deviation from the power-law
    behaviour is observed. The deviations from the power-law regime in
    both cases are not due to finite size effects, as can be seen from a systematic finite size study. The solid line represents the power law obtained for observational data.  Results indicate that the turbulent flow and a reconnection ruled by twisting are the ingredients that control energy statistics.}\label{fig3}
\end{figure}

It is interesting to investigate the role of different
ingredients of the theoretical model on the statistical properties of
energy released and waiting times, in order to identify the main triggering mechanisms for the occurrence of solar flares.  We start by considering that, instead of being driven by the turbulent flow, the magnetic flux tubes might move along purely random trajectories and the cumulative twist is calculated by
assigning a random vorticity at each footprint. Results in
Figure~\ref{fig3} show that the suppression of the turbulent flow leads
to an energy distribution that is exponential rather than a power
law. Next, we consider the case where there is a single magnetic flux
tube evolving in the turbulent flow, eliminating the possible role of
interactions among tubes. We observe in Figure~\ref{fig3} that the
power-law regime in the peak energy distribution is recovered. These two
results strongly suggest that the ingredient responsible for the
power-law in the energy distribution is the turbulent motion of the
footprints anchored into the photosphere, and not tube-tube
interactions. We finally consider the case of several tubes having
different degree of interaction, i.e. either interacting ($\lambda_{\rm R} <
1$ and $\lambda_{\rm R} > 1$) or non-interacting ($\lambda_{\rm R} = 1$) tubes.  Results shown in Figure~\ref{fig3} confirm that interactions do not modify 
the distribution of solar flare energies. Interestingly, models with and without avalanching exhibit the same scaling exponent for the peak flare energy distribution, suggesting that indeed small flares share similar statistical properties with major flares.

\begin{figure}
  \centering
  \includegraphics[width=\columnwidth]{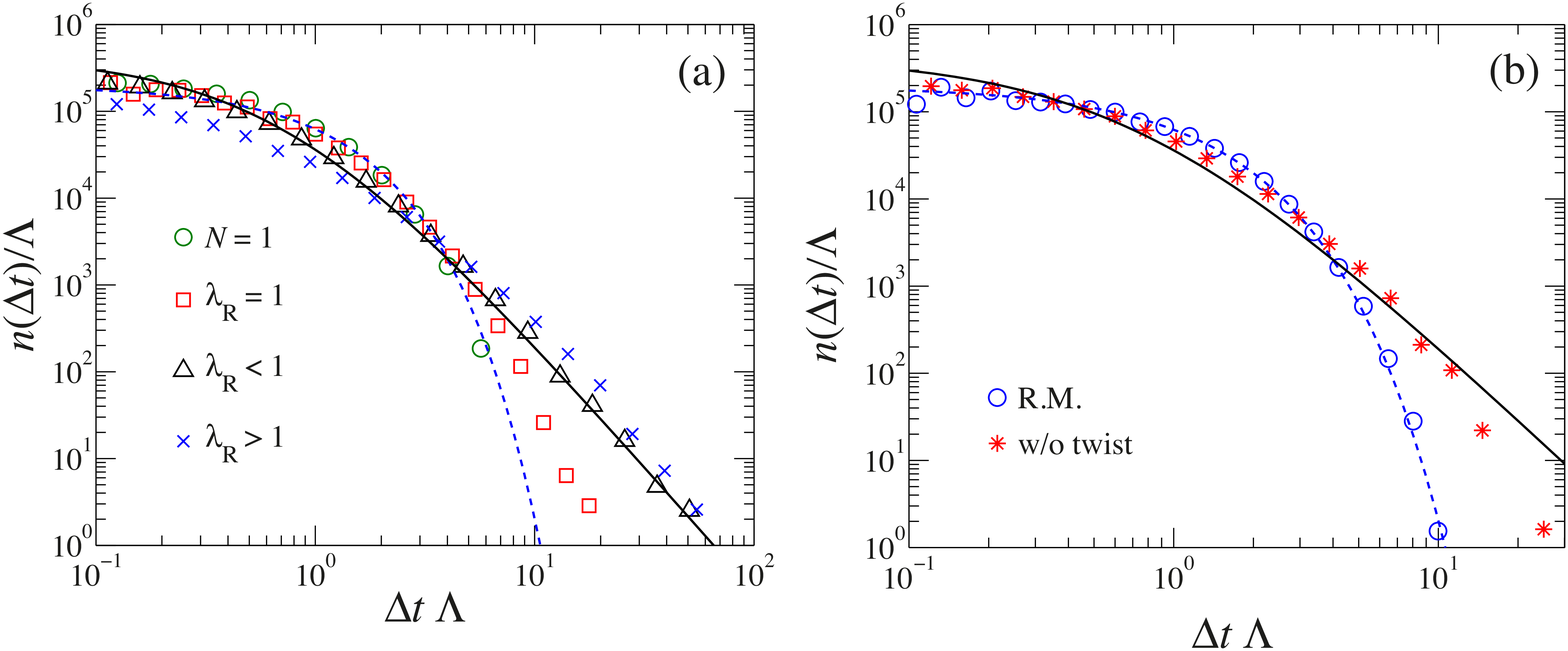}
  \caption{{\bf Physical ingredients leading to flare organisation in
      time}.  Different cases are compared to
    extract the crucial ingredients of the observed waiting time
    distribution. The system size is $L = 512$ for all curves. (a) Waiting time distribution for simulations where a single flux
    tube ($N = 1$) evolves in the turbulent flow and reconnects at the
    twisting threshold. The distribution exhibits an exponential decay
    (dashed blue line).  Conversely, for different degrees of
    interaction (interacting tubes, $\lambda_{\rm R} < 1$ and $\lambda_{\rm R} > 1$) and non-interacting tubes ($\lambda_{\rm R} = 1$), one recovers the behaviour measured in Figure~\ref{fig4}. (b) Distributions evaluated for the evolution of $N = 400$ tubes in the two cases: Interacting tubes ($\lambda_{\rm R} < 1$), whose footprints diffuse by random motion (denoted by ``R.M.'') and
    evolution in the turbulent flow {\color{black} w/o twist}. In both
    cases deviations from the observational result are observed (the dash blue line denotes an exponential decay). The solid line in (a) and (b) represents the best fit for the largest system size, $L = 2048$. Results indicate that tube interactions rule the temporal organisation of flares.
  }\label{fig5}
\end{figure}

We now consider the waiting time distribution for the previous
cases. Indeed, in Figure~\ref{fig5}, we see that the case of random
footprint motion and the case of a solitary tube are well described by
a Poissonian distribution (dashed line). This implies that, although
essential to the recovery of a scale-free energy distribution, the
turbulent fluid flow alone is not able to provide the right temporal
organization of solar flare occurrence. If more tubes are considered,
the distribution starts to deviate from a Poissonian one. For
coexisting but non-interacting tubes ($N > 1$, $\lambda_{\rm R} =1$), the
turbulent flow in the photosphere is able to induce time correlations
between them, although not sufficiently to reproduce the observational
results. Indeed, the physical correlations are fully recovered only
for interacting tubes ($N > 1$, $\lambda_{\rm R} <1$ and $\lambda_{\rm R} > 1$).  From our results, we can conclude that, whereas the turbulent photospheric flow is the main
mechanism responsible for the energy distribution, the interaction
between magnetic tubes is what introduces the right temporal
correlations in the process.

\subsection*{Time-energy Correlations}

We further investigate the statistical features of time-energy correlations\cite{lucilla2}.  For each catalogue, we analyse how the flare energies are organised in time, by evaluating the probability that a flare with energy $E_i$ is followed by a flare with energy larger than $\lambda E_i$ under the
conditions that their temporal distance $\Delta t$ is smaller than a
certain threshold $t_{\rm th}$, $P(\lambda | t_{\rm th})= P(E_{i+1}/E_{i}
>\lambda | \Delta t_i < t_{\rm th})$.  For each catalogue, this probability
fluctuates wildly due to statistical noise.  Therefore to eliminate
this noise, we evaluate the same probability also in a
synthetic catalogue generated by reshuffling the flare energies with
respect to their occurrence time, such that energy and time are
uncorrelated by construction.  We then consider the
difference between the conditional probabilities, $\delta P(\lambda |
t_{\rm th})$, evaluated in the two data sets. This difference is different from zero only if
significant time-energy correlations are present in the original
catalogue.  In particular, if $|\delta P(\lambda | t_{\rm th})|$ is larger
than zero, it is more likely to find two consecutive flares satisfying
both conditions ($E_{i+1}/E_{i} >\lambda$ and $\Delta t_i < t_{\rm th}$)
in the real rather than in the reshuffled catalogue (see Methods).  By
using the same technique we also compute the conditional probability
difference $\delta P(E_{\rm th} | t_{\rm th})$ to observe a flare energy
larger than a given threshold $E_{\rm th}$ after an waiting time smaller
than $t_{\rm th}$.  We consider the behaviour of both conditional
probability differences for a range of parameters $\lambda$, $ t_{\rm th}$
and $ E_{\rm th}$.

\begin{figure}
  \centering
  \includegraphics[width=0.5\columnwidth]{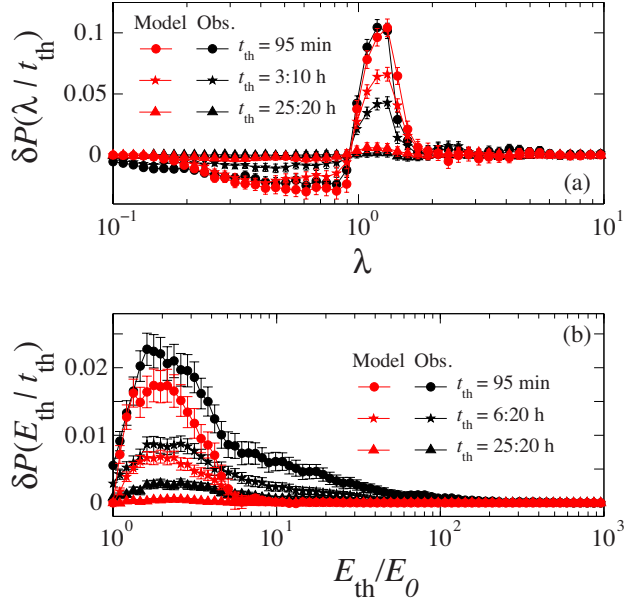}
  \caption{{\bf Flare energies and occurrence times are correlated}.
    (a) The conditional probability difference $\delta P(\lambda |
    t_{\rm th})=\delta P(E_{i+1}/E_{i} >\lambda | \Delta t_i < t_{\rm th})$ is
    plotted as function of the parameter $\lambda$ for different
    $t_{\rm th}$.  Black and red colours represent the observational and
    numerical catalogues, respectively. Error bars report the standard
    deviation of the probability evaluated for the reshuffled catalogue
    (see Methods). $\delta P$ is different from zero beyond error bars
    for a range of $\lambda >1$ indicating that, both in the
    observational and numerical catalogue, for consecutive flares
    occurring within $3$ hours the energy of the second flare is
    larger than the energy of the previous flare.  (b) The conditional
    probability difference $\delta P (E_{\rm th} | t_{\rm th})=\delta
    P(E_{i+1}>E_{\rm th} | \Delta t_i < t_{\rm th})$ as function of the energy
    threshold $E_{\rm th}$ for different $t_{\rm th}$. In order to compare the
    different catalogues, energies are normalized by $E_0 = 1.5\times
    10^{-6}$ W m$^{-2}$ for observational data and $E_0 =150$ for
    numerical data. $\delta P$ is different from zero beyond error
    bars for energy thresholds $E_{\rm th}$ up to about $10 E_0$
    indicating that, both in the observational and numerical catalogue,
    as consecutive flares become more distant in time, the probability
    to find the following flare with energy higher than $E_{\rm th}$
    becomes smaller. To convert the numerical units of time, $\Delta t_{\rm num}$, into physical units, $\Delta t$, we have used the expression $\Delta t_{\rm num} = \Lambda \Delta t / \Lambda_n$, where $\Lambda_n$ and $\Lambda$ are the inverse of the average waiting times for the numerical and observational catalogues, respectively.} \label{fig8}
\end{figure}

In Figure~\ref{fig8} we see that the probability differences are very
well described by our model with $\lambda_{\rm R} < 1$.
In particular, for both, numerical and
observational results, $\delta P(\lambda | t_{\rm th})$ is different from
zero beyond error bars. This implies that it is very likely that for
close-in-time flares the second one will have slightly larger energy
than the previous one (the maximum is for $\lambda \gtrsim 1$), as far
as their separation in time is shorter than approximately $25$
hours. These energy correlations decrease as the temporal separation increases. Conversely, it is very unlikely to observe in real catalogues close-in-time flares where the second one has a smaller energy ($\delta P<0$ for $(\lambda<1$)). Furthermore, in Figure~\ref{fig8}b curves for $\delta P(E_{\rm th}|t_{\rm th})$ are different from zero beyond error bars and
decrease with increasing $t_{\rm th}$.  This implies that the probability
to find couples of successive flares with the second flare having
energy higher than $E_{\rm th}$ decreases, if one includes events separated by a
larger $\Delta t$ in the analysis. For large $E_{\rm th}$ numerical
results deviate from the observational ones. This can be due to the
finite size of our simulation, which  imposes an upper limit to the flux tube sizes, and therefore limits the maximum energy of flares (see also Figure~\ref{fig2}, where finite-size effects in the
energy distribution are analysed). Note that the agreement between
observational and numerical data is very good, suggesting that, in fact,
the turbulent flow and magnetic flux tube interactions 
induce correlations between energy and time in the occurrence of solar flares.

\begin{figure}
  \centering
  \includegraphics[width=0.5\columnwidth]{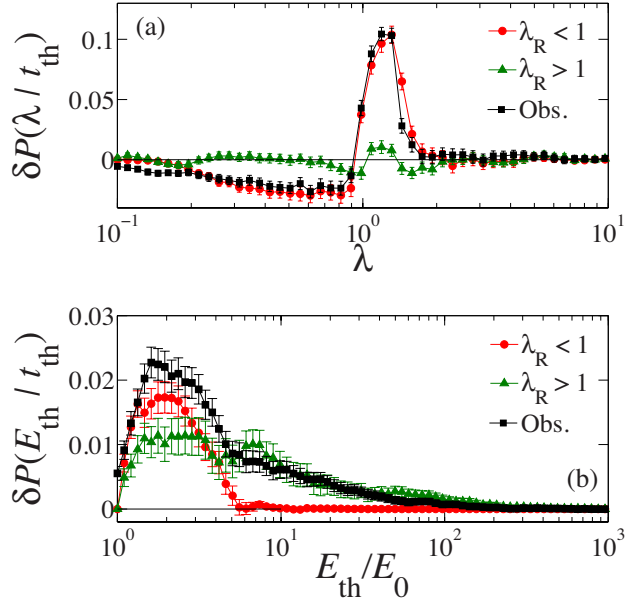}
  \caption{{\bf Physical ingredients leading to time-energy correlations}.
    (a) The conditional probability difference $\delta P(\lambda |
    t_{\rm th})$ is plotted as function of the parameter $\lambda$ for
    $t_{\rm th} = 95$ min. Error bars report the standard deviation of the probability evaluated for the reshuffled catalogue (see Methods). For the case $\lambda_{\rm R} < 1$, $\delta P$ is different from zero beyond error bars and in quantitative agreement with observational data. Conversely, for the case $\lambda_{\rm R} > 1$, $\delta P>0$ in a narrow range, with an amplitude in disagreement with correlations measured in catalogues. (b) The conditional probability difference $\delta P (E_{\rm th} | t_{\rm th})$ as function of the energy threshold $E_{\rm th}$ for both interacting models, $\lambda_{\rm R} > 1$ and $\lambda_{\rm R} < 1$. In order to compare the different catalogues, energies are normalised by $E_0 =150$ for both catalogues. $\delta P$ is different from zero beyond error bars for energy thresholds $E_{\rm th}$ up to about $10 E_0$ (for $\lambda_{\rm R} < 1$) and $100 E_0$ (for $\lambda_{\rm R} > 1$) indicating that, both models can reproduce this kind of correlations. For computing the conditional probabilities, we have chosen $t_{\rm th} = 95$ min. By increasing $t_{\rm th}$, we recover, for both catalogues, that the time-energy correlations decrease as in Figure~\ref{fig8}.} \label{fig9m}
\end{figure}

It is important to notice that agreement with observational data is only obtained for interacting tubes ($\lambda_{\rm R} < 1$), not for non-interacting tubes, single tube, and random motion of tube footprints, where $\delta P(E_{\rm th} | t_{\rm th})
\simeq 0$ and $\delta P(\lambda | t_{\rm th}) \simeq 0$ are found. For the case $\lambda_{\rm R} > 1$, we do not obtain full quantitative agreement with time-energy correlations measured in experimental catalogues (see Figure~\ref{fig9m}). Data still predict that the next flare statistically has slightly larger energy than the previous one but in a narrower $\lambda$ range and with a smaller probability. Moreover, the case $\lambda_{\rm R} > 1$ is not able to reproduce the anti-correlations for $\lambda < 1$. These results suggest that the relaxation mechanism corresponding to $\lambda_{\rm R} < 1$ seems to be more appropriate to reproduce time-energy interaction, as compared to the avalanche process characterised by $\lambda_{\rm R} > 1$.

Finally, we have performed extensive simulations varying every parameter of the model, namely, the critical twist $\Phi_{\rm c}$, the magnetic flux tube density via $N$, and the size ratio of the magnetic flux tubes $r_{\rm c}/R$ (where $r_{\rm c}$ and $R$ are the inner and outer radii of the magnetic flux tubes). For all cases, results for flare statistics and time-energy correlations remain unchanged as shown in the {\color{black} Supplementary
Figures 3, 4, and 5}.

\section*{Discussion}

It is well accepted in the literature that solar flare occurrence is a
process driven by magnetic reconnection. Since high
quality satellite data became accessible, several studies have evidenced that the
statistics of this phenomenon is complex: It exhibits scale-free
energy distribution and a nontrivial waiting time distribution.  A
number of theoretical models attempted to reproduce such statistics
with different approaches.  This study implements magnetic
reconnection in a model framework that enables us to test the role of
the different physical ingredients on observed statistical patterns.
In particular, we have shown that the energy distribution are ruled by the turbulent features of the flow in the photospheric plasma.

More precisely, that, if tube footprints simply diffuse
in the corona in absence of turbulent flow, the observed distribution
of flare energies would be exponential, namely it would exhibit a
characteristic flare size. Moreover, the turbulent flow alone is not
sufficient to fully reproduce the statistical patterns of real
data. Indeed the evolution of a single tube or several non-interacting
tubes in the corona, exhibiting the observed energy distribution, is
not sufficient to account for temporal correlations.

The detailed analysis of the energy organisation in time indicates
that turbulence and tube interactions are the essential physical
ingredients controlling solar flare occurrence. Proving time-energy correlations is the first step towards any forecasting model. This could be formalised by implementing the phenomenological laws, as done for earthquakes \cite{Ogata} (ETAS model), and would open a novel field of investigation.

\section*{METHODS}
\subsection*{Evaluation of the turbulent flow}

For modelling the two-dimensional turbulent flow, we have used a
two-dimensional lattice Boltzmann model of size $L$ with a cell
configuration $D2Q9$ ($2$ dimensions and $9$ discrete velocity
vectors)\cite{LBE1}. In order to induce turbulence, we have included the following forcing term in the 2D Navier-Stokes equation
\begin{equation}
  {\bf F} = A_0 \left (\sin(k_{\rm f} x) \cos(k_{\rm f} y), - \cos(k_{\rm f} x) \sin(k_{\rm f} 
    y)\right ) ,
\end{equation}
where $A_0$ is a constant, $k_{\rm f} = 2\pi q$, and $q$ varies in time such that each value of $q$ is used during an interval of time $s = (4 q/L)^{-\mu}$, and then is increased by one. As an initial value, we take $q=2$. Here, $\mu$ is a tuning parameter to control the spectrum of the energy of the turbulent flow (in our simulations, $\mu = 5/3$). The coefficient $4$ defines the minimum wave number (due to space discretisation limitations). This forcing term satisfies the incompressibility condition $\nabla \cdot {\bf F} = 0$. The kinematic viscosity of the fluid is set to $\nu = 10^{-3}$ and the forcing coefficient to $A_0 =
10^{-8}$. As initial conditions for the fluid, we choose density $\rho = 1$, and velocity ${\bf u} = (0,0)$. Furthermore, we also impose a large scale dissipation mechanism to avoid vortex condensation\cite{2Dturbulence}. All the values are in numerical units. 

Once the fluid has reached the turbulent regime, we insert $N$ magnetic flux tubes in random positions. The position of the respective footprints for each tube $l$, denoted by ${\bf x}_{l+}$ for the positive footprint and
${\bf x}_{l-}$ for the negative one, is a function of time and evolves
as,
\begin{equation}\label{motion:eq}
  {\bf x}_{l\pm}(t+\delta t) = {\bf x}_{l \pm}(t) + 
  {\bf u}({\bf x}_{l\pm}) \delta t  ,
\end{equation}
where ${\bf u}({\bf x})$ is the velocity of the fluid at position
${\bf x}$. 

Then, we can define $w_{l +}$ and $w_{l -}$ as the cumulative twist in
the positive and negative footprint, respectively, evolving according to the equation,
\begin{equation}\label{vorticity:eq}
  w_{l\pm}(t+\delta t) = w_{l \pm}(t) + (\nabla \times {\bf u})_z \delta t 
\quad .
\end{equation}
Note that the component used to twist the magnetic flux tubes is the z-component of the vorticity, since the velocity lies on the two-dimensional space.
 
As initial conditions, the flux tubes have an outer radius of $R = 4$ cells and zero twist, $w_{l\pm} = 0$. If the positive footprint of a tube comes very close to its negative partner (less than two lattice nodes), we reset the tube to the initial condition (initial length and zero twist) and relocate it at a random position inside the simulated zone. The vanishing of these tubes can be seen as small reconnection processes with negligible released energy. Because of space discretisation in our numerical simulations, the position of each footprint is, in general, not located at a fluid grid point, therefore we use bilinear
interpolation to calculate the velocity at the footprint position.
Note that the motion of the footprints, see Eq.~\eqref{motion:eq}, as well as the twist, see Eq.~\eqref{vorticity:eq}, are additive (See Supplementary Figure 6), in the sense that they systematically inject electric currents and associated magnetic energy in the system.

The magnetic field lines are modelled as lines wrapped around semi-circular flux tubes, forming, when twisted, a spring-shaped bundle (see Figure~\ref{fig1}). Therefore, we can assume that the total energy of a tube is given by the length of the magnetic line, which depends on the twisting and the size of the semi-circular tube. Thus, when a flux tube is not twisted, its energy equals  $E_l = \pi R$ (here and throughout the following calculation, we have omitted the proportionality constant to get the right units of energy). On the other hand, if a flux tube is twisted, the spring-shaped bundle can be parametrised by
\begin{eqnarray}
\begin{aligned}
  {\bf R}_l(\omega) = ([(R+r_{\rm c}) &+ r_{\rm c} \cos(\omega)] \cos(\xi), r_{\rm c} \sin(\omega), \\ &[(R+r_{\rm c}) + r_{\rm c} \cos(\omega)] \sin(\xi)),
  \end{aligned}
\end{eqnarray}
where $r_{\rm c}$ is the cross section radius of the semi-circular tube. The value $\xi$ depends on $\omega$ as follows: $\xi = \Theta \omega$, where $\Theta$ is a constant that controls the number of turns that the magnetic line makes around the semi-circular tube. In this coordinate system $(x,y,z)$, the photosphere is located at the plane $x-y$. The length of this parametric curve is given by the integral,
\begin{equation}\label{integral}
  E_l = \int_0^\pi \sqrt{\frac{d{\bf R}_l}{d\omega}\cdot \frac{d{\bf R}_l}{d\omega}}\; d\omega \quad .
\end{equation}

According to the kink instability theory \cite{hood2}, the twist is defined as $\Phi = \pi R B_{\theta}/r_{\rm c} B_z$, where $B_{\theta}$ and $B_z$ are the tangential and perpendicular components of the magnetic field to the plane $x-y$ at the footprint ($\omega = 0$). Therefore, the ratio $B_{\theta}/B_z$ is equivalent to the ratio between the $y$ and $z$ components of the derivative of the parametric curve, $B_{\theta}/B_z = (dR_{ly}/d\omega)/(dR_{lz}/d\omega)$ with $\omega = 0$, and we can conclude that $\Phi = \pi R \Theta/(2 r_{\rm c} + R)$ is the twist. In our model, $\Theta$ denotes the cumulative total angle due to the vorticity of the fluid, $\Theta = w_{l+} + w_{l-}$.

The integral in Eq.~\eqref{integral} does not possess an analytical solution. However, we can assume that $r_{\rm c} \ll R$, obtaining a very simple expression:
\begin{eqnarray}\label{energy_tube}
    E_l = \pi R \quad .
\end{eqnarray}
{\color{black} Note that this expression is identical to the expression for an untwisted flux tube for any values of $r_{\rm c}$ and $R$.}

Once a magnetic flux tube reaches the critical twist $\Phi_{\rm c}$, the tube releases its entire energy and vanishes. In order to keep the tube density in a stationary state and produce a sufficient statistics, a new tube is placed at a random position inside the simulated zone with the initial condition ($w_{l\pm} = 0$ and $R = 4$ cells). When several magnetic flux tubes reach the critical twist within the same temporal interval $\delta t=1$ , we sum the energies of the tubes into a single event. For the case $\lambda_{\rm R} < 1$, we have also evaluated the distributions keeping simultaneous events separate (see Supplementary Figure 5) and verified that the main properties of solar flare statistics remain unchanged. For the case of $\lambda_{\rm R} > 1$, since flaring occurs through an avalanching process, we perform the measurement of events as follows. Once a flare occurs, we stop the fluid and observe if it triggers other flares. We measure the peak flux energy as the largest flare that occurs within the avalanche process, and the total energy as the sum of all flares. The duration of flares is taken as the total number of released flares. Afterwards, the dynamics of the photospheric flow is restarted. Our model cannot accommodate helicity conservation during the magnetic reconnection process \cite{berger1999}. Therefore, it is assumed that each helical kink instability ejects the unstable flux tube out of the simulation volume to infinity (physically, that would mean that each flare is eruptive). However, it seems that this effect is not relevant to reproduce the solar flare statistics.

Note that the energy stored by a tube scales with the tube length and therefore has an upper cutoff controlled by the system size. We have also run the simulations for different initial conditions, finding that our statistical results remain unchanged. In particular, for the cases where $r_{\rm c}/R < 1$ (but not necessarily $r_{\rm c}/R \ll 1$), we have solved numerically the integral in Eq.~\eqref{integral} considering terms up to order $(r_{\rm c}/R)^{10}$.

We have implemented our numerical code using CUDA C. The simulations for a fixed set of parameters run three weeks on a cluster of $12$ graphic cards, Nvidia Tesla M2075, each one containing $448$ GPU
cores.

\subsection*{Conditional probability analysis}

Each flare $i$ in the numerical and observational catalogues is
characterized by its starting time $t_i$ and its peak-flux energy
$E_i$. From each catalogue we evaluate the conditional probability $
P(\lambda | t_{\rm th})= P(E_{i+1}/E_{i} > \lambda | \Delta t_i < t_{\rm th})$
to find the energy of the next flare ($E_{i+1}$) being larger than
$\lambda$ times the energy of the previous flare ($E_i$), if their
temporal distance, $\Delta t \equiv t_{i+1} - t_{i}$, is smaller than
a certain threshold, $t_{\rm th}$. For comparison, the same conditional
probability is evaluated from a reshuffled sequence of the same energy
time series. In such synthetic catalogues we expect that flare
energies and occurrence times are totally uncorrelated.  Keeping
$\lambda$ and $ t_{\rm th}$ fixed, we compute the quantity $ P^*(\lambda |
t_{\rm th})$ for $10^5$ independent realisations of the reshuffled
catalogue, obtaining an ensemble of values which follows a Gaussian
distribution with mean value $ Q(\lambda | t_{\rm th})$ and standard
deviation $\sigma (\lambda | t_{\rm th})$.  We then define $\delta
P(\lambda | t_{\rm th}) \equiv P(\lambda | t_{\rm th})- Q(\lambda | t_{\rm th})$.
If the absolute value $|\delta P(\lambda | t_{\rm th})| > \sigma(\lambda | t_{\rm th})$, a significant difference in the number of pairs of sequential energies $(E_{i}, E_{i+1})$ satisfying both conditions exists between
the real and the reshuffled catalogue. By using the same technique we
also compute the conditional probability difference $\delta P(E_{\rm th} |
t_{\rm th}) \equiv P(E_{i+1} > E_{\rm th} | \Delta t_i < t_{\rm th}) - Q(E_{\rm th},
t_{\rm th})$.

\bibliography{report}

\begin{acknowledgments}
  We acknowledge financial support from the European Research Council
  (ERC) Advanced Grant 319968-FlowCCS, and the Brazilian agencies
  CNPq, CAPES and FUNCAP. MM and AK would also like to acknowledge
  many fruitful discussions with Fabrizio Lombardi and Micha Wasem.
\end{acknowledgments}

\section*{Author contributions}
All authors conceived and designed the research, analyzed the data,
worked out the theory, and wrote the manuscript.

\section*{Additional information}
\textbf{Competing financial interests:} The authors declare no competing
financial interests.

\section*{Figure Legends}
\pagebreak
\newpage

\end{document}